\documentclass[aps,prl, reprint,superscriptaddress]{revtex4-2}

\usepackage{amsmath}
\usepackage{here}
\usepackage[symbol]{footmisc}
\usepackage[dvipdfmx]{}
\usepackage{comment}
\usepackage{graphicx}
\usepackage{dcolumn}
\usepackage{bm}
\newcommand{\degr}{^{\circ}}

\begin{document}

\title{Gamma-ray Observation of the Cygnus Region in the 100 TeV Energy Region}

\author{M.~Amenomori}
\affiliation{Department of Physics, Hirosaki University, Hirosaki 036-8561, Japan} 
\author{Y.~W.~Bao}
\affiliation{School of Astronomy and Space Science, Nanjing University, Nanjing 210093, China}
\author{X.~J.~Bi}
\affiliation{Key Laboratory of Particle Astrophysics, Institute of High Energy Physics, Chinese Academy of Sciences, Beijing 100049, China}
\author{D.~Chen}
\altaffiliation{chending@bao.ac.cn}
\affiliation{National Astronomical Observatories, Chinese Academy of Sciences, Beijing 100012, China}
\author{T.~L.~Chen}
\affiliation{Department of Mathematics and Physics, Tibet University, Lhasa 850000, China}
\author{W.~Y.~Chen}
\affiliation{Key Laboratory of Particle Astrophysics, Institute of High Energy Physics, Chinese Academy of Sciences, Beijing 100049, China}
\author{Xu~Chen}
\affiliation{Key Laboratory of Particle Astrophysics, Institute of High Energy Physics, Chinese Academy of Sciences, Beijing 100049, China}
\author{Y.~Chen}
\affiliation{School of Astronomy and Space Science, Nanjing University, Nanjing 210093, China}
\author{Cirennima}
\affiliation{Department of Mathematics and Physics, Tibet University, Lhasa 850000, China}
\author{S.~W.~Cui}
\affiliation{Department of Physics, Hebei Normal University, Shijiazhuang 050016, China}
% ...
\collaboration{The Tibet AS$\gamma$ Collaboration}

\date{\today}

\begin{abstract}
We report observations of gamma-ray emissions with energies in the 100~TeV energy region from the Cygnus region in our Galaxy.
Two sources are significantly detected in the directions of the Cygnus OB1 and OB2 associations.
Based on their positional coincidences, we associate one with a pulsar PSR~J2032$+$4127 and the other
mainly with a pulsar wind nebula PWN~G75.2+0.1 with the pulsar moving away from its original birthplace situated around the centroid of the observed gamma-ray emission.
This work would stimulate further studies of particle acceleration mechanisms at these gamma-ray sources.
\end{abstract}

\maketitle

\section{Introduction}\label{intro}
The Cygnus region is a large active star-forming complex in our Galaxy, hosting numerous pulsar wind nebulae (PWNe) and supernova remnants (SNRs) as well as Wolf-Rayet stars, OB associations, open clusters etc.
This region is a natural laboratory to study cosmic-ray acceleration and transport, and has been observed in various wavebands.
Especially interesting in light of TeV gamma-ray observations are the Cygnus OB1 and OB2 associations.

In the direction of Cygnus OB2, the HEGRA Imaging Air Cherenkov Telescope (IACT) discovered a gamma-ray source TeV~J2032$+$4130 \cite{HEGRA2002, HEGRA2005}, and 
follow-up observations have been made by other IACTs including Whipple \cite{Whipple2007}, MAGIC \cite{MAGIC2008} 
and VERITAS \cite{VERITASJ2031, VERITAS2018}.
They reported gamma-ray fluxes of roughly 3\% that of the Crab nebula above 1~TeV and source extensions of radii 0.1$\degr$\textendash 0.2$\degr$.
Meanwhile, air-shower arrays generally reported fluxes much higher than those of IACTs at multi-TeV energies \cite{Milagro2007a, Milagro2012, ARGO2012, ARGO2014, HAWC2017}. 
For example, ARGO \cite{ARGO2012} employed a larger variable-size window with a radius of approximately 1.5$^{\circ}$ at 1~TeV and reported flux data points 
5 \textendash 10 times those of IACTs at multi-TeV energies, and 
the flux reported at 7~TeV by HAWC with a window of radius 0.7$^{\circ}$ was approximately twice that based on their point-source analysis \cite{HAWC2017}, which 
appears to be consistent with the IACT measurements.  
A recent HAWC paper \cite{HAWC2021} reported that there is a largely extended `Cocoon' region (HAWC~J2030$+$409) with a Gaussian width of $2.13^{\circ}$,
counterpart of the GeV Cygnus Cocoon \cite{Fermi2011}, underneath the gamma-ray emission of HAWC~J2031$+$415 with a width of $0.27^{\circ}$, 
counterpart of TeV~J2032$+$4130.
A plausible explanation of the discrepancy in the measured fluxes between IACTs and air-shower arrays is that
IACTs extract the emission of HAWC~J2031$+$415 by counting the extended emission of the Cocoon region in estimating the background, while
air-shower arrays tend to integrate part of the gamma-ray emission of the Cocoon region inside their relatively large window sizes as well as the gamma-ray 
emission of HAWC~J2031$+$415. 
Additionally, HAWC detected eHWC~J2030$+$412 above 56~TeV with an extension of 0.18$^{\circ}$ \cite{HAWC2020PRL}, 
although the association of eHWC~J2030$+$412 with TeV~J2032$+$4130/PSR~J2032$+$4127 is a matter of debate, given that eHWC~J2030$+$412 is 
$0.36^{\circ}$ away from TeV~J2032$+$4130 and $0.33^{\circ}$ away from PSR~J2032$+$4127. 

Located 0.18$^{\circ}$ southeast of the centroid of TeV~J2032$+$4130 is 
the radio pulsar PSR~J2032$+$4127, with the distance from the Earth of 1.4\textendash1.7 kpc \cite{OB21, OB22, OB23},
a characteristic age of 180~kyr, and a spin-down luminosity of $1.7 \times 10^{35}$~erg/s \cite{PSROB21}.
PSR~J2032$+$4127 forms a binary system with a massive B0Ve star MT91~213, and 
a flux enhancement of this binary system in the 2017 autumn periastron period was observed at the 0.2 \textendash 2~TeV energy region by VERITAS and MAGIC
and at the 0.3 \textendash 10~keV energy region by {\it Swift}-X-Ray Telescope and {\it NuSTAR} \cite{BinaryX, Binary, Binary2020}.


In the direction of Cygnus OB1, a gamma-ray source MGRO~J2019$+$37 was discovered by Milagro at the median energy of 12~TeV with a source extension of $0.3\degr$ \cite{Milagro2007b}, and this source was also recently observed by HAWC 
\cite{HAWC2020PRL, HAWC2020}.
VERITAS observed the same region above 0.6~TeV and separated the gamma-ray emissions into two sources: VER~J2019$+$368 and 
VER~J2016$+$371 \cite{VERITASJ2019}.
VERITAS also reported that the morphology of VER~J2019$+$368 is asymmetrical with $\sigma_{\mathrm{major}} (\sigma_{\mathrm{minor}})~\approx~0.34\degr~(0.13\degr)$.
The flux of VER~J2019$+$368 measured by VERITAS in 2014 \cite{VERITASJ2019} is a few times higher than that in 2018 \cite{VERITAS2018} due to 
different collection areas of photon integration windows.

Located 0.36$\degr$ east of the centroid of VER~J2019$+$368 is the radio pulsar PSR~J2021$+$3651, 
having a characteristic age of 17~kyr and a spin-down luminosity of $3.4 \times 10^{36}$~erg/s \cite{PSROB11}.
PSR~J2021$+$3651 is one of the relatively few pulsars which have gamma-ray pulsations observed by {\it Fermi}-LAT \cite{D2}.
The distance $D$ to the pulsar has remained a matter of debate. 
$D \sim 12$~kpc was suggested by the dispersion measure \cite{PSROB11} and $D \geq 5$~kpc by the pulsar polarization rotation measure \cite{D2}.
Comparing the HI column density along the pulsar line of sight with the hydrogen absorbing column density implied $D \sim 10$~kpc \cite{Hessels2004} or 
$3-4$~kpc \cite{D1}, and $D~=~1.8^{+1.7}_{-1.4}$~kpc was obtained from the absorption-distance relation using red-clump stars in the direction of the pulsar \cite{Kiri}. 
This value is consistent with 1.58~kpc, which is the estimated distance to Cygnus OB1 \cite{D5,D4}.
PSR~J2021$+$3651 appears to form a pulsar wind nebula referred to as PWN~G75.2$+$0.1 \cite{Hessels2004} and 
the radio and X-ray morphologies of this nebula feature a bright bow-shaped tail extending westwards from the pulsar, indicating 
that the pulsar is moving eastwards with its birthplace as far west as the apparent end of the tail at $\approx 0.2\degr$ west of the current pulsar 
position \cite{VERITASJ2019, Mizuno2017}.
In addition, a number of hard X-ray sources were identified along the western edge of VER~J2019$+$368 by $NuSTAR$ space observations 
including a young massive stellar cluster \cite{NuSTAR}.

\section{Experiment and Data Analysis}
The Tibet air-shower (AS) array has been observing cosmic rays and gamma rays above TeV energies since 1990 at Yangbajing (90.522$\degr$E, 30.102$\degr$N; 4300~m above sea level) in Tibet, China \cite{Tibet1992}.
In this work, we use data obtained by the Tibet air shower array combined with the muon detector array during 719 live days from 2014 February to 2017 May.
Our data analysis method and data selection criteria as well as the array configuration are the same as described in our previous papers \cite{Tibet2019, Tibet2021}.
We use the Equi-Zenith-Angle method \cite{Tibet2005}
to estimate the gamma-ray excess count and the number of background events.
Twenty OFF regions are taken, and 
the radius of the analysis window is variable depending on the recorded air-shower size from approximately 0.7$^{\circ}$ at lower energies around 
10 TeV to a lower limit of 0.5$^{\circ}$.

In the following section, we discuss two gamma-ray sources detected significantly above 10~TeV in the directions of Cygnus OB1 and OB2 respectively.

\section{Results and Discussions}
\subsection*{Cygnus OB2}
\begin{figure}
  \begin{center}
	  \includegraphics[width=1\columnwidth]{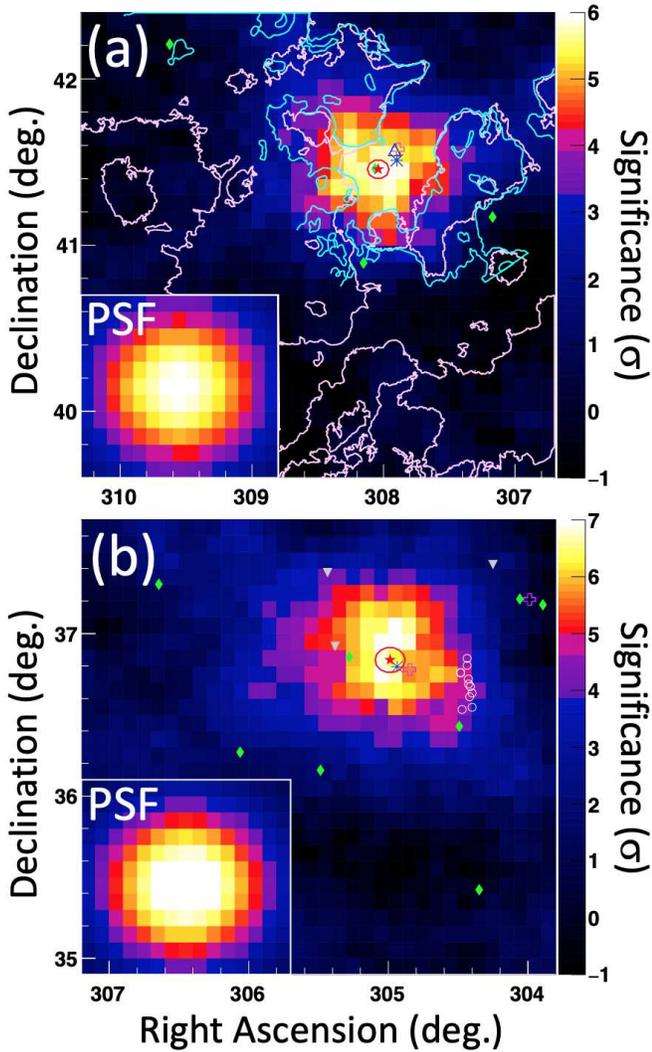}
  \end{center}
  \caption{Significance maps around the two gamma-ray emission sources detected above 10~TeV in the directions of Cygnus OB2 (a) and OB1 (b), smoothed by search windows (see the text). 
The point spread function (PSF) is shown in the inset figure. 
The red filled star with a position error circle is the centroid of 
TASG J2032$+$414/TASG J2019$+$368 obtained by this work,
while the magenta open cross is the centroid 
of VER J2031$+$415 \cite{Binary}/VER J2019$+$368 \cite{VERITAS2018}
and the blue asterisk is that of 
HAWC J2031$+$415 \cite{HAWC2021}/3HWC J2019$+$367 \cite{HAWC2020}.
The green filled diamonds show {\it Fermi}-LAT sources \cite{3FGL}.
(a) The blue open triangle indicates the centroid of MAGIC J2031$+$4134 \cite{Binary}.
The green filled diamond coincident with our gamma-ray emission centroid is the pulsar PSR~J2032$+$4127.
The sky-blue contours indicate 1420~MHz radio emissions provided by the Canadian Galactic Plane Survey \cite{CGPS}, and
the pink contours indicate 24~$\mu$m infrared emissions by the Cygnus-X {\it Spitzer} Legacy Survey \cite{MIPS1, MIPS2}.
(b) The white open circles are {\it NuSTAR} X-ray sources \cite{NuSTAR}, and
the gray filled inverted triangles are Wolf-Rayet stars \cite{WRcat}.  
The green filled diamond located at 0.23$\degr$ east of our emission centroid is the pulsar PSR~J2021$+$3651. 
The magenta open cross located at (R.A., Dec.) = (303.99$\degr$, 37.21$\degr$) is another VERITAS source VER~J2016$+$371 \cite{VERITAS2018}, which is not detected significantly in this work.
}
  \label{2Dmap}
\end{figure}

Figure~\ref{2Dmap}(a) shows a detection significance map around the gamma-ray source detected by this work with photon energies above 10~TeV in the direction of 
Cygnus OB2.  
The sky is gridded in $0.1\degr \times 0.1\degr$ pixels and the significance value of each pixel calculated according to \cite{LiMa}
is smoothed by a circular search window of radius $R_{\mathrm{w}}$ centered at the pixel. 
Assuming a symmetrical 2D Gaussian distribution for the gamma-ray excess, we fit the events within the $4\degr \times 4\degr$ region around the source using
the unbinned maximum likelihood method. 
The centroid of gamma-ray emissions detected at the pre-trial (post-trial) detection significance of 5.3$\sigma$ (4.7$\sigma$) above 10~TeV is estimated at (R.A., Dec.) = ($308.04\degr \pm 0.08\degr, 41.46\degr \pm 0.06\degr$).
We name this source TASG~J2032$+414$.
The location of TASG~J2032$+414$ is in good agreement with that of the pulsar PSR~J2032$+$4127 and consistent with that of HAWC~J2031$+$415 \cite{HAWC2021}
at the 1.7$\sigma$ level, while it appears to deviate from that of TeV~J2032$+$4130 reported in \cite{Binary} at the 2.8$\sigma$ level.
We also find that most of the gamma-ray emission detected above 10~TeV is confined inside a void 
where the radio (frequency 1420~MHz) \cite{CGPS} and infrared (wavelength 24~$\mu$m) \cite{MIPS1, MIPS2} emissions are very weak.
This morphology was also seen by VERITAS \cite{VERITAS2018}.

\begin{figure}
  \begin{center}
	  \includegraphics[width=1\columnwidth]{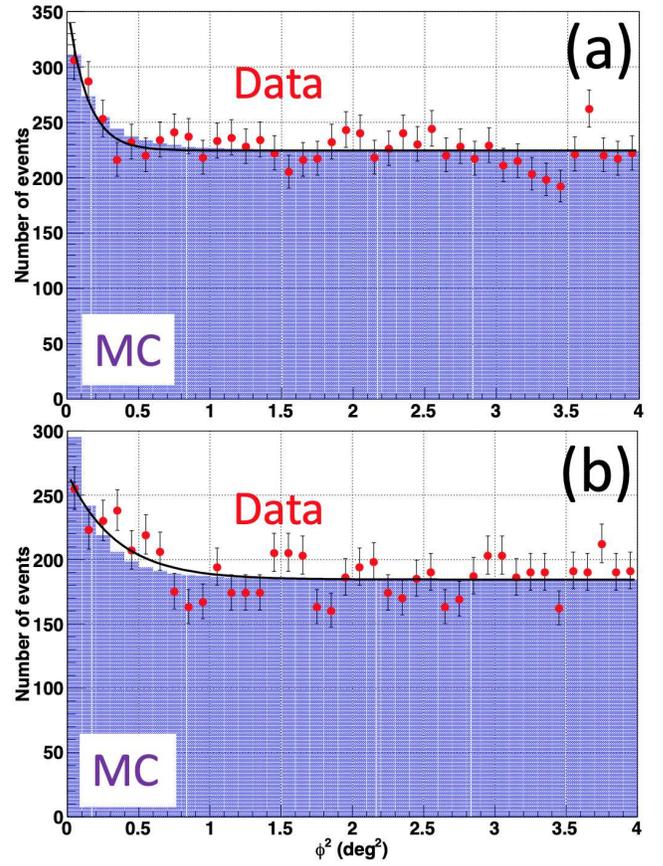}
  \end{center}
  \caption{Number of events observed with photon energies above 10~TeV as a function of the square of the opening angle between the estimated arrival direction and the centroid of 
TASG J2032$+$414 in (a) and TASG J2019$+$368 in (b). 
The red filled circles are the experimental data, with the best-fit Gaussian function indicated by the solid line. The blue histogram is 
  the distribution of events expected by the MC simulation assuming a point-like gamma-ray source. }
  \label{PHI2}
\end{figure}

Figure~\ref{PHI2}(a) shows the distribution of the number of observed events above 10 TeV as a
function of the square of the opening angle $\phi$ between the estimated arrival direction and the TASG~J2032$+414$ centroid.
To estimate a possible source extension, 
we perform the $\chi^2$ fitting of the data with the function $A\exp[-\phi^2 / 2(\sigma_{\mathrm{PSF}}^2 + \sigma_{\mathrm{EXT}}^2)] + N_{\mathrm{BG}}$
where $A$ and $\sigma_{\mathrm{EXT}}$ are two fitting parameters and $\sigma_{\mathrm{PSF}} = 0.36\degr$ and $N_{\mathrm{BG}} = 224.5$ are
the point spread function (PSF) of our instrument above 10~TeV and the number of background events estimated from the background cosmic-ray data, 
we get $\sigma_{\mathrm{EXT}} = 0.00^{\circ} \pm 0.14^{\circ}$, which is consistent with 
that obtained from the maximum likelihood fitting described above.
The $\chi^2$/ndf of the fitting is 33.8/38. 
With a large error of 0.14$^{\circ}$, the $\sigma_{\mathrm{EXT}}$ value above 10~TeV does not indicate whether TASG~J2032$+414$ is extended or not 
even though it is consistent with the previous measurements at multi-TeV energies by IACTs, ARGO and HAWC within the 2$\sigma$ level 
\cite{HEGRA2005, Whipple2007, MAGIC2008, ARGO2012, VERITAS2018, Binary, HAWC2021}. 

At higher energies above 40~TeV, our source location is consistent with that of PSR~J2032$+$4127, 
and the estimated value of $\sigma_{\mathrm{EXT}} = 0.16\degr \pm 0.09\degr$ suggests a slight source extension at the 1.9$\sigma$ level
(refer to Supplemental Material Figure~S1 and S2).
There seems to be a tension  
between the centroid of our source above 40~TeV and that of eHWC~J2030$+$412 above 56~TeV \cite{HAWC2020PRL} at a statistical significance of 3.5$\sigma$, 
which might result from the complex morphology of the Cygnus Cocoon region as these source locations are obtained  
under the assumption of a Gaussian spatial distribution for gamma-ray signals.

\begin{figure}
  \begin{center}
	  \includegraphics[width=1\columnwidth]{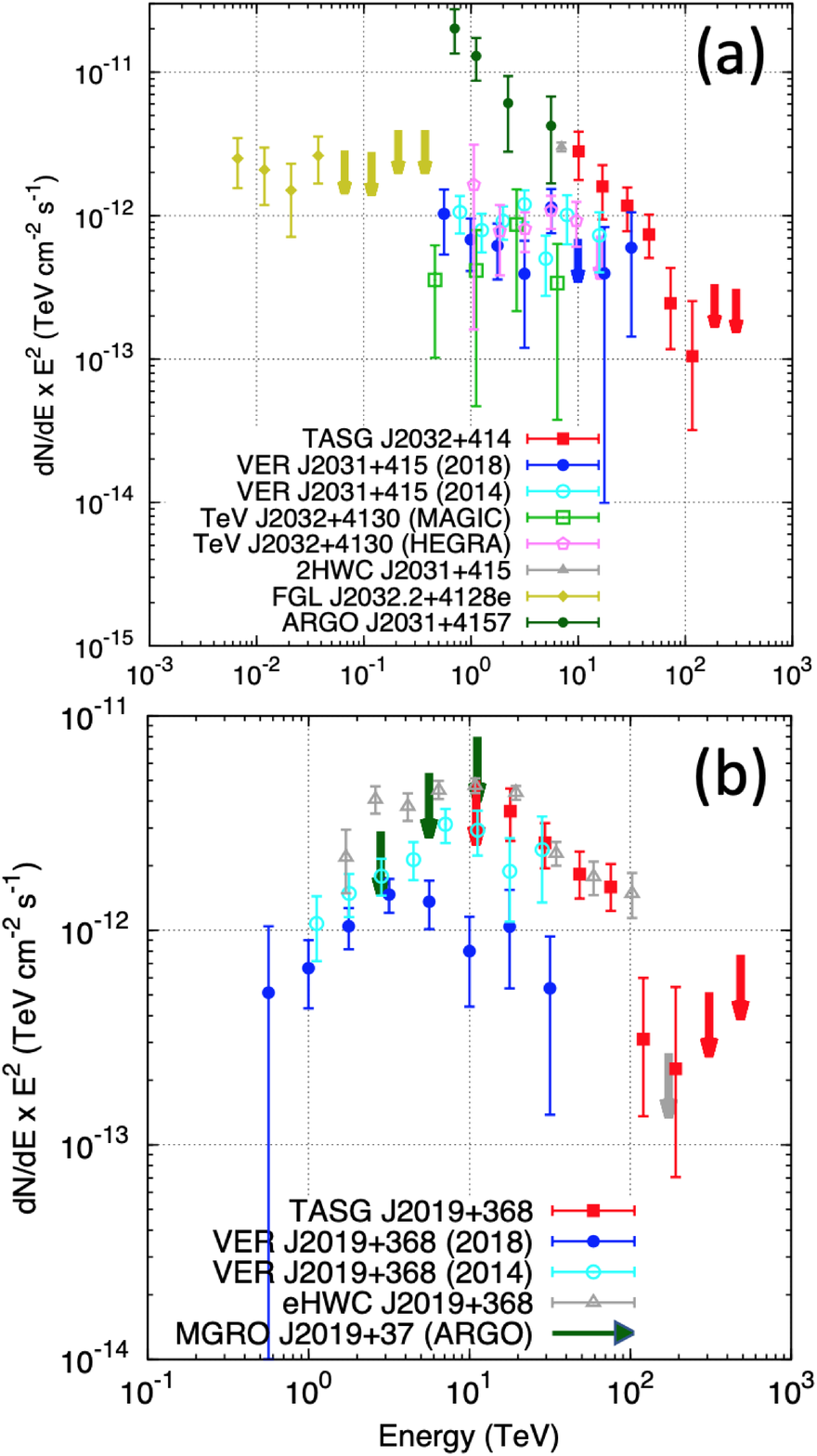}
  \end{center}
  \caption{Differential gamma-ray energy spectra of (a) TASG~J2032$+414$ 
  and (b) TASG~J2019$+$368 
with 95\% C.L. upper limits measured by this work (red filled squares/arrows).
In both panels, the blue filled circles/arrow (sky-blue open circles) show the gamma-ray spectrum of reported by VERITAS in 2018 (2014) \cite{VERITAS2018, VERITASJ2031, VERITASJ2019}, 
the gray open triangles/arrow by HAWC \cite{HAWC2017, HAWC2020PRL}, and the dark-green filled 
circles/arrows by ARGO \cite{ARGO2012}. Additionally in panel (a),  
the gold filled diamonds are reported by {\it Fermi}-LAT \cite{VERITAS2018}, the green open squares by MAGIC \cite{MAGIC2008}, and the pink pentagons/arrow by HEGRA \cite{HEGRA2005}.
The upper limits of {\it Fermi}-LAT, HAWC and VERITAS are at the 95\% confidence level, while
those of HEGRA in (a) and ARGO in (b) are at the 99\% and 90\% confidence levels, respectively. 
}
  \label{FLUX}
\end{figure}

Figure~\ref{FLUX}(a) shows the differential energy spectrum of TASG~J2032$+414$ (red filled squares and downward arrows).
Although there is a discrepancy in flux at multi-TeV energies as explained in the Introduction, 
our flux data points above 10~TeV are consistent with previous measurements of IACTs when the spill-over of gamma-ray signals outside their integration radius is taken 
into account.
Our spectrum from 10~TeV to 120~TeV can be expressed by a simple power-law as $dF / dE = N_0 (E/40~\mathrm{\mbox{TeV}})^{-\Gamma}$
where $N_0 = (4.13 \pm 0.83) \times 10^{-16}$ TeV$^{-1}$cm$^{-2}$s$^{-1}$ is the differential gamma-ray flux at 40~TeV 
and $\Gamma = 3.12 \pm 0.21$ is the spectral index ($\chi^2$/ndf = 1.6/4).

Given that the TASG~J2032$+414$ centroid is in good agreement with the location of the radio pulsar PSR~J2032$+$4127, 
the observed gamma rays would be produced by relativistic electrons injected by the pulsar through the inverse-Compton scattering (ICS) 
with synchrotron and ambient photons.
Unfortunately, the time span of our data does not cover the 2017 autumn periastron period of the binary system PSR~J2032$+$4127/MT91~213.
TASG~J2032$+414$, therefore, is considered to be associated with the pulsar itself rather than the whole binary system.
VERITAS and MAGIC detected TeV gamma-ray emissions coincident with TeV~J2032$+$4130, which is considered 
to be the pulsar wind nebula of PSR~J2032$+$4127
with its centroid 0.2$^{\circ}$ away from the location of PSR~J2032$+$4127.
If the pulsar is moving in the southeast direction along the elongation of the TeV gamma-ray emission as suggested in \cite{VERITASJ2031},
it would be natural to consider that gamma rays above 10~TeV produced by parent electrons with energies of $\gtrsim$ 100~TeV are 
confined around the current pulsar location, 
while parent electrons of TeV gamma rays were accelerated earlier when the pulsar was at the centroid of TeV~J2032$+$4130. 

The other two TeV sources reported earlier in this region, VER~J2019$+$407/3HWC~J2020$+$403 and 2HWC~J2024$+$417$^{\ast}$
are not significantly detected in this work.
We obtain 99\% C.L. integral flux upper limits above 10~TeV (100~TeV) of
$1.0 \times 10^{-13}$~cm$^{-2}$s$^{-1}$ ($1.7 \times 10^{-15}$~cm$^{-2}$s$^{-1}$)
for VER~J2019$+$407/3HWC~J2020$+$403, and 
$0.92 \times 10^{-13}$~cm$^{-2}$s$^{-1}$ ($2.0 \times 10^{-15}$~cm$^{-2}$s$^{-1}$)
for 2HWC~J2024$+$417$^{\ast}$.

\subsection*{Cygnus OB1}
Figure~\ref{2Dmap}(b) shows a significance map above 10~TeV in the direction of Cygnus OB1 obtained by this work.
The centroid of gamma-ray emissions is estimated at (R.A., Dec.) = ($304.99\degr \pm 0.11\degr, 36.84\degr \pm 0.08\degr$) 
with the pre-trial (post-trial) detection significance of 6.7$\sigma$ (6.2$\sigma$). 
We name this source TASG~J2019$+$368.
The centroid of TASG~J2019$+$368 is consistent with that reported by HAWC \cite{HAWC2020} within the $1\sigma$ level, 
and by VERITAS \cite{VERITAS2018} within the 2$\sigma$ level. 
Possible particle acceleration sites are also indicated in the figure: 
{\it NuSTAR} X-ray sources \cite{NuSTAR}, Wolf-Rayet stars \cite{WRcat} and {\it Fermi}-LAT sources \cite{3FGL}. 
The pulsar PSR~J2021$+$3651, located 0.23$^{\circ}$ east of the TASG~J2019$+$368 centroid, has a nebula extending westwards 
from the pulsar, PWN ~G75.2$+$0.1, which is coincident with the location of TASG~J2019$+$368.

Figure~\ref{PHI2}(b) shows the $\phi^2$ distribution of the events observed above 10~TeV.
The experimental data can be fitted with a Gaussian function 
with a source extension of $\sigma_{\mathrm{EXT}} = 0.28\degr~\pm~0.07\degr$ above 10~TeV, consistent with the extension reported by VERITAS \cite{VERITAS2018} 
(HAWC \cite{HAWC2020PRL}) at the 2.1$\sigma$ (0.3$\sigma$) level.
The $\chi^2$/ndf of the fitting is 49.1/38. 

At higher energies above 40~TeV, 
our source location is consistent with that of VERITAS \cite{VERITAS2018} and HAWC \cite{HAWC2020, HAWC2020PRL},  
and the estimated value of $\sigma_{\mathrm{EXT}} = 0.22\degr \pm 0.05\degr$ suggests that the extension could become smaller 
as photon energy increases (refer to Supplemental Material Figure~S1 and S2). 
This tendency was also seen earlier by HAWC \cite{HAWC2020PRL} 
($\sigma_{\mathrm{EXT}} = 0.30\degr \pm 0.02\degr (0.20\degr \pm 0.05\degr)$ above $\sim$1~TeV (56~TeV)).

Figure~3(b) shows the differential gamma-ray energy spectrum of TASG~J2019$+$368, which is 
in good agreement with the HAWC spectrum and consistently connects with the VERITAS spectrum reported in 2014.
Our spectrum can be expressed either as $dF / dE = N_0 (E/40~\mathrm{\mbox{TeV}})^{-\Gamma}$
with $N_0 = (10.6 \pm 1.3) \times 10^{-16}$ TeV$^{-1}$cm$^{-2}$s$^{-1}$ and $\Gamma = 2.70 \pm 0.13$ ($\chi^2$/ndf = 10.4/5), 
or including an exponential cutoff as $dF / dE = N_0 (E/40~\mathrm{\mbox{TeV}})^{-\Gamma} \exp(-E/E_{\mathrm{cut}})$ with
$N_0 = (3.6 \pm 2.0) \times 10^{-15}$ TeV$^{-1}$cm$^{-2}$s$^{-1}$, $\Gamma = 1.6 \pm 0.5$ and $E_{\mathrm{cut}} = 44 \pm 21$~TeV ($\chi^2$/ndf = 3.0/4). 
The latter fitting is preferred because the best-fit power-law function of the former fitting conflicts with the HAWC upper limit at 170~TeV.


The current position of the pulsar PSR~J2021$+$3651 is 0.23$^{\circ}$ east of our gamma-ray emission centroid, and 
the estimated location of the pulsar's birthplace resides in our 1$\sigma$ position error circle, 
indicating that the gamma-ray emission observed above 10~TeV by this work would be mainly caused by 
relativistic electrons produced by PSR~J2021$+$3651 around the time when the pulsar was very young.
If the pulsar birthplace is the gamma-ray emission centroid obtained by this work,
the transverse velocity of the pulsar can be expressed as $v = 420~(D/1.8~\mathrm{kpc})~(\tau/17~\mathrm{kyr})^{-1}$~km/s, where $\tau$ is the pulsar age.
Among known pulsar velocities \cite{velo}, $v = 420$~km/s is a plausible value. 
We consider that the observed gamma-ray emission is mainly associated with PSR~J2021$+$3651, although there might be 
some contributions from nearby sources such as the $NuSTAR$ sources.

The other TeV source VER~J2016$+$371 reported earlier in this region is not significantly detected by this work.
We obtain a 99\% C.L. integral flux upper limit above 10~TeV (100~TeV) of
$0.91 \times 10^{-13}$~cm$^{-2}$s$^{-1}$ ($1.4 \times 10^{-15}$~cm$^{-2}$s$^{-1}$) for VER~J2016+371.

\section{Conclusions}
Using the Tibet air shower array combined with the underground muon detector array, 
we have detected two gamma-ray sources in the direction of the Cygnus region significantly with photon energies above 10~TeV: 
TASG~J2032$+$414 in the direction of Cygnus OB2 at (R.A., Dec.) = ($308.04\degr \pm 0.08\degr, 41.46\degr \pm 0.06\degr$), coincident with PSR~J2032$+$4127, and
TASG~J2019$+$368  in the direction of Cygnus OB1 at (R.A., Dec.) = ($304.99\degr \pm 0.11\degr, 36.84\degr \pm 0.08\degr$), 
coincident with PWN~G75.2$+$0.1.

We provide for the first time flux data points from 40~TeV to 120~TeV for the gamma-ray emission region in the direction of Cygnus OB2, 
and we confirm the HAWC spectrum up to 100~TeV and provide flux data points from 120~TeV to 200~TeV for the gamma-ray emission region in the direction of Cygnus OB1.  
For both cases, the observed gamma-ray emissions likely result from relativistic electrons injected by the pulsar through the ICS with synchrotron and ambient photons.
PSR~J2021$+$3651 could be moving eastwards with a velocity of $\sim 400$~km/s away from its original birthplace situated around the observed gamma-ray emission centroid. 
Future observations and theoretical studies would
shed more light on the physical mechanisms of particle acceleration at these two high-energy gamma-ray sources.

\begin{acknowledgments}
The collaborative experiment of the Tibet Air Shower Arrays has been 
conducted under the auspices of the Ministry of Science and Technology
of China and the Ministry of Foreign Affairs of Japan. 
This work was supported in part by a Grant-in-Aid for Scientific Research on Priority Areas from the Ministry of Education, Culture, Sports, Science and Technology, 
and by Grants-in-Aid for Science Research from the Japan Society for the Promotion of Science in Japan.
This work is supported by the National Key R\&D Program of China (No. 2016YFE0125500),
the Grants from the National Natural Science Foundation of China (Nos. 11533007, 11673041, 11873065, 11773019, 11773014, 11633007, 11803011, 
and 11851305), and the Key Laboratory of Particle Astrophysics, Institute of High Energy Physics, CAS.
The research presented in this paper has used data supplied through the Canadian Galactic Plane Survey.
This work is also supported by the joint research program of the Institute for Cosmic Ray Research (ICRR), the University of Tokyo.
\end{acknowledgments}



\end{document}